\documentclass[seceqn]{elsart}
\usepackage{epsfig,amsmath}
\newcommand{\Lmu}{L}

\begin{document}

\begin{flushright}
\newlength{\prepw}
\settowidth{\prepw}{Alberta Thy 00-00}
\begin{minipage}{\prepw}
TTP09-41\\
SFB/CPP-09-110\\
Alberta Thy 16-09
\end{minipage}
\end{flushright}

\begin{frontmatter}
\title{Matching QCD and HQET heavy--light currents at three loops}
\author[Karlsruhe]{S.~Bekavac},
\author[Karlsruhe,Novosibirsk]{A.G.~Grozin},
\author[Karlsruhe]{P.~Marquard},
\author[Edmonton]{J.H.~Piclum},
\author[Edmonton]{D.~Seidel},
\author[Karlsruhe]{M.~Steinhauser}
\address[Karlsruhe]{Institut f\"ur Theoretische Teilchenphysik,
\mbox{Karlsruhe Institute of Technology (KIT)}, 76128 Karlsruhe, Germany}
\address[Novosibirsk]{Budker Institute of Nuclear Physics,
Novosibirsk 630090, Russia}
\address[Edmonton]{Department of Physics, University of Alberta,
Edmonton, Alberta T6G 2G7, Canada}

\begin{abstract}
We consider the currents formed by a heavy and a light quark within
  Quantum Chromodynamics and compute the matching to Heavy Quark Effective
  Theory to three-loop accuracy. As an application we obtain the
  third-order perturbative corrections to ratios
  of $B$-meson decay constants.
\end{abstract}
\end{frontmatter}

\section{Introduction}
\label{S:Intro}

Quite often there are problems within Quantum Chromodynamics (QCD)
involving a single heavy quark with momentum
\begin{equation}
  p = m v + k\,,
  \label{Intro:p}
\end{equation}
where $m$ is the on-shell heavy-quark mass and $v^2=1$.
In situations when the characteristic residual momentum is small
($|k^\mu|\ll m$)
and light quarks and gluons have small momenta ($|k_i^\mu|\ll m$)
it is possible to use a simpler effective field theory ---
Heavy Quark Effective Theory (HQET) (see, e.g.,
the review~\cite{N:94} and the textbooks~\cite{MW:00,G:04}).
Its Lagrangian is a series in $1/m$.
QCD operators are also given by series in $1/m$
in terms of HQET operators.
Here we shall consider $\overline{\mbox{MS}}$ renormalized
heavy--light QCD quark currents
\begin{equation}
j(\mu) = Z_j^{-1}(\mu) j_0\,,\quad
j_0 = \bar{q}_0 \Gamma Q_0\,,
\label{Intro:j}
\end{equation}
where $\Gamma$ is a Dirac matrix,
and the index 0 means bare quantities.
They can be expressed via operators in HQET
\begin{equation}
j(\mu) = C_\Gamma(\mu) \tilde{\jmath}(\mu)
+ \frac{1}{2m} \sum_i B_i(\mu) O_i(\mu)
+ \mathcal{O}\left(\frac{1}{m^2}\right)\,,
\label{Intro:match}
\end{equation}
where
\begin{equation}
\tilde{\jmath}(\mu) = \tilde{Z}_j^{-1}(\mu) \tilde{\jmath}_0\,,\quad
\tilde{\jmath}_0 = \bar{q}_0 \Gamma Q_{v0}\,,
\label{Intro:jtilde}
\end{equation}
$Q_{v0}$ is the bare HQET heavy-quark field
satisfying $\rlap/v Q_{v0} = Q_{v0}$,
and $O_i$ are dimension-4 HQET operators
with appropriate quantum numbers.
We shall not discuss $1/m$ corrections in this paper;
our main subject is the matching coefficients $C_\Gamma(\mu)$.

The coefficients $C_\Gamma(\mu)$ have been
calculated at one-loop order in the pioneering paper~\cite{EH:90}.
At two loops, they were calculated in Ref.~\cite{BG:95},
and corrected in Ref.~\cite{G:98}.\footnote{The results incorporating this correction
  are given by formulae~(5.65--5.68) and Table~5.1 in~\cite{G:04}.
  Note a misprint in this table: in the row for $\Gamma=\gamma^1$,
  the term with $C_F$ in the square bracket,
  $-1453/48$, should be $+1453/48$.}
All-order results in the large-$\beta_0$ limit
were obtained in~\cite{BG:95} (see also~\cite{NS:95}).
Asymptotics of perturbative series were investigated
in a model-independent way in Ref.~\cite{CGM:03}.

In the present paper we calculate the matching coefficients $C_\Gamma$
up to three loops.
These coefficients are useful for obtaining matrix elements
of QCD currents (such as $f_B$) from results of lattice HQET simulations
(see, e.g., Refs.~\cite{lattice:09,Sommer})
or HQET sum rules (see, e.g., Refs.~\cite{SR,N:94}).


\section{Matching}
\label{S:Match}

There are eight Dirac structures giving non-vanishing quark currents
in four dimensions:
\begin{eqnarray}
\Gamma &{}={}& 1\,,\quad
\rlap/v\,,\quad
\gamma_\bot^\alpha\,,\quad
\gamma_\bot^\alpha \rlap/v\,,
\nonumber\\
&&\gamma_\bot^{[\alpha} \gamma_\bot^{\beta]}\,,\quad
\gamma_\bot^{[\alpha} \gamma_\bot^{\beta]} \rlap/v\,,\quad
\gamma_\bot^{[\alpha} \gamma_\bot^{\beta} \gamma_\bot^{\gamma]}\,,\quad
\gamma_\bot^{[\alpha} \gamma_\bot^{\beta} \gamma_\bot^{\gamma]} \rlap/v\,,
\label{Match:Gamma}
\end{eqnarray}
where $\gamma_\bot^\alpha = \gamma^\alpha - \rlap/v v^\alpha$,
and square brackets mean antisymmetrization.
The last four of them can be obtained from the first four
by multiplying by the 't~Hooft--Veltman $\gamma_5^{\mbox{\scriptsize HV}}$.
We are concerned with flavour non-singlet currents only,
therefore, we may also use the anticommuting $\gamma_5^{\mbox{\scriptsize AC}}$
(there is no anomaly).
The currents renormalized at a scale $\mu$
with different prescriptions for $\gamma_5$ 
are related by~\cite{LV:91}
\begin{eqnarray}
\left(\bar{q} \gamma_5^{\mbox{\scriptsize AC}} Q\right)_\mu &{}={}&
Z_P(\mu) \left(\bar{q} \gamma_5^{\mbox{\scriptsize HV}} Q\right)_\mu\,,
\nonumber\\
\left(\bar{q} \gamma_5^{\mbox{\scriptsize AC}} \gamma^\alpha Q\right)_\mu &{}={}&
Z_A(\mu) \left(\bar{q} \gamma_5^{\mbox{\scriptsize HV}} \gamma^\alpha Q\right)_\mu\,,
\nonumber\\
\left(\bar{q} \gamma_5^{\mbox{\scriptsize AC}} \gamma^{[\alpha} \gamma^{\beta]} Q\right)_\mu &{}={}&
Z_T(\mu) \left(\bar{q} \gamma_5^{\mbox{\scriptsize HV}} \gamma^{[\alpha}
  \gamma^{\beta]} Q\right)_\mu\,,
\label{Match:Larin}
\end{eqnarray}
where the finite renormalization constants $Z_{P,A,T}$
can be reconstructed from the differences of the anomalous dimensions
of the currents.
Multiplying $\Gamma$ by $\gamma_5^{\mbox{\scriptsize AC}}$
does not change the anomalous dimension.
In the case of $\Gamma=\gamma^{[\alpha} \gamma^{\beta]}$,
multiplying it by $\gamma_5^{\mbox{\scriptsize HV}}$
just permutes its components,
and also does not change the anomalous dimension,
therefore,
\begin{equation}
Z_T(\mu) = 1\,;
\label{Match:ZT}
\end{equation}
$Z_{P,A}(\mu)$ are known up to three loops~\cite{LV:91}.

The anomalous dimension of the HQET current~(\ref{Intro:jtilde})
does not depend on the Dirac structure $\Gamma$.
Therefore, there are no factors similar to $Z_{P,A}$ in HQET.
Multiplying $\Gamma$ by $\gamma_5^{\mbox{\scriptsize AC}}$
does not change the matching coefficient.
Therefore, the matching coefficients for the currents
in the second row of~(\ref{Match:Gamma})
are not independent:
they can be obtained from those for the first row.
In the $v$ rest frame, where $\rlap/v=\gamma^0$, we have
\begin{eqnarray}
Z_P(\mu) &{}={}&
\frac{C_{\gamma_5^{\mbox{\scriptsize AC}}}(\mu)}%
{C_{\gamma_5^{\mbox{\scriptsize HV}}}(\mu)} =
\frac{C_1(\mu)}%
{C_{\gamma^0 \gamma^1 \gamma^2 \gamma^3}(\mu)}\,,
\nonumber\\
Z_A(\mu) &{}={}&
\frac{C_{\gamma_5^{\mbox{\scriptsize AC}} \gamma^0}(\mu)}%
{C_{\gamma_5^{\mbox{\scriptsize HV}} \gamma^0}(\mu)} =
\frac{C_{\gamma^0}(\mu)}%
{C_{\gamma^1 \gamma^2 \gamma^3}(\mu)}
\nonumber\\
&{}={}&
\frac{C_{\gamma_5^{\mbox{\scriptsize AC}} \gamma^3}(\mu)}%
{C_{\gamma_5^{\mbox{\scriptsize HV}} \gamma^3}(\mu)} =
\frac{C_{\gamma^3}(\mu)}%
{C_{\gamma^0 \gamma^1 \gamma^2}(\mu)}\,,
\nonumber\\
Z_T(\mu) &{}={}&
\frac{C_{\gamma_5^{\mbox{\scriptsize AC}} \gamma^0 \gamma^1}(\mu)}%
{C_{\gamma_5^{\mbox{\scriptsize HV}} \gamma^0 \gamma^1}(\mu)} =
\frac{C_{\gamma^0 \gamma^1}(\mu)}%
{C_{\gamma^2 \gamma^3}(\mu)}
\nonumber\\
&{}={}&
\frac{C_{\gamma_5^{\mbox{\scriptsize AC}} \gamma^2 \gamma^3}(\mu)}%
{C_{\gamma_5^{\mbox{\scriptsize HV}} \gamma^2 \gamma^3}(\mu)} =
\frac{C_{\gamma^2 \gamma^3}(\mu)}%
{C_{\gamma^0 \gamma^1}(\mu)} = 1\,.
\label{Match:Ratios}
\end{eqnarray}
In particular, two matching coefficients are equal:
\begin{equation}
C_{\gamma_\bot \rlap{\scriptsize/}v}(\mu) =
C_{\gamma_\bot^{[\alpha} \gamma_\bot^{\beta]}}(\mu)\,.
\label{Match:CT}
\end{equation}
In the following we shall consider only the matching coefficients
for the first four Dirac structures in~(\ref{Match:Gamma}).

In order to find the matching coefficients $C_\Gamma(\mu)$,
we equate on-shell matrix elements of
the left- and right-hand side of Eq.~(\ref{Intro:match}).
They are obtained by considering transitions
of the heavy quark with momentum $p=mv+k$~(\ref{Intro:p})
to the light quark with momentum $k_q$:
\begin{equation}
{\langle}q(k_q)|j(\mu)|Q(mv+k){\rangle} =
C_\Gamma(\mu) {\langle}q(k_q)|\tilde{\jmath}(\mu)|Q_v(k){\rangle}
+ \mathcal{O}\left(\frac{k}{m},\frac{k_q}{m}\right)\,.
\label{Match:match}
\end{equation}
The on-shell matrix elements are\footnote{Both on-shell matrix elements of the
  renormalized currents are UV-finite. Both contain IR divergences, which are
  the same on the left- and right-hand sides 
  of~(\ref{Match:match}), yielding a finite $C_\Gamma(\mu)$.}
\begin{eqnarray}
{\langle}q(k_q)|j(\mu)|Q(p){\rangle} &{}={}&
\bar{u}_q(k_q) \Gamma(p,k_q) u(p)\,
Z_j^{-1}(\mu) Z_Q^{1/2} Z_q^{1/2}\,,
\nonumber\\
{\langle}q(k_q)|\tilde{\jmath}(\mu)|Q_v(k){\rangle} &{}={}&
\bar{u}_q(k_q) \tilde{\Gamma}(k,k_q) u_v(k)\,
\tilde{Z}_j^{-1}(\mu) \tilde{Z}_Q^{1/2} \tilde{Z}_q^{1/2}\,,
\label{Match:onshell}
\end{eqnarray}
where $\Gamma(p,k_q)$ and $\tilde{\Gamma}(k,k_q)$ are the bare vertex functions,
$Z_Q$ and $Z_q$ are the on-shell wave-function renormalization constants
of the heavy and the light quark in QCD,
$\tilde{Z}_Q$ is the on-shell wave-function renormalization constant
of the HQET quark field $Q_v$,
and $\tilde{Z}_q$ differs from $Z_q$ because there are no $Q$ loops in HQET.
The difference between $u(mv+k)$ and $u_v(k)$ is of order $k/m$,
and can be neglected.
It is most convenient to use $k=k_q=0$,
then the $\mathcal{O}(1/m)$ term is absent.
The QCD vertex has two Dirac structures:
\[ \Gamma(mv,0) = \Gamma \cdot (A + B \rlap/v)\,. \]
This leads to
\[
\bar{u}(0) \Gamma(mv,0) u(mv) = \bar{\Gamma}(mv,0)\,\bar{u}(0) \Gamma u(mv)
\quad\mbox{with}\quad
\bar{\Gamma}(mv,0) = A + B\,.
\]
The scalar vertex function $\bar{\Gamma}(mv,0)$ can be obtained by
multiplying the diagrams by a projector and taking the trace.
For the first two Dirac structures in~(\ref{Match:Gamma}),
the projector $(\rlap/v+1)$ can be used;
for the next two structures, $(\rlap/v+1)\gamma_\alpha$.
The HQET vertex has just one Dirac structure.
Therefore,
\begin{equation}
C_\Gamma(\mu) =
\frac{\bar{\Gamma}(mv,0) Z_j^{-1}(\mu) Z_Q^{1/2} Z_q^{1/2}}%
{\tilde{\Gamma}(0,0) \tilde{Z}_j^{-1}(\mu) \tilde{Z}_Q^{1/2} \tilde{Z}_q^{1/2}}\,.
\label{Match:main}
\end{equation}
Here $m$ is the on-shell mass of the heavy quark
(because the external heavy quark with $p^2=m^2$
should be on its mass shell).
Therefore, mass-counterterm vertices have to be taken into account
on all $Q$ lines.

If all flavours except $Q$ are massless,
all loop corrections to $\tilde{\Gamma}(0,0)$,
$\tilde{Z}_Q$, and $\tilde{Z}_q$ contain no scale
and hence vanish: $\tilde{\Gamma}(0,0)=1$,
$\tilde{Z}_Q=1$, $\tilde{Z}_q=1$.
If there is another massive flavour (charm in $b$-quark HQET),
this is no longer so.
The vertex $\tilde{\Gamma}(0,0)$ and $\tilde{Z}_Q$
have been calculated up to three loops in Ref.~\cite{GSS:06}.
The massless-quark on-shell wave-function renormalization constant
in the presence of a massive flavour has been calculated
up to three loops in Ref.~\cite{CKS:98}
(an explicit expression can be found in Appendix~A of Ref.~\cite{GSS:06}).
The three-loop renormalization of the HQET current $\tilde{Z}_j$
has been calculated in Ref.~\cite{CG:03}.

If all flavours except $Q$ are massless,
$\Gamma(mv,0)$, $Z_Q$, and $Z_q$ contain a single scale $m$.
The on-shell heavy-quark wave-function renormalization constant $Z_Q$
has been calculated at three loops in~\cite{MR:00}
(and confirmed in Ref.~\cite{MMPS:07});
$Z_q$ has been found in Ref.~\cite{CKS:98}.
The vertex $\Gamma(mv,0)$ is the subject
of the present paper.
If there is another flavour with a non-zero mass $m_c<m$,
there are two scales, and calculations become more difficult.
The renormalization constant $Z_Q$ has been calculated
in this case, up to three loops, in Ref.~\cite{BGSS:07}
(the master integrals appearing in this case
are discussed in Ref.~\cite{BGSS:08}).
The vertex $\Gamma(mv,0)$ and $Z_q$ with two masses
are considered in Sect.~\ref{S:Calc} and~\ref{S:Zq}.
The three-loop renormalization of the QCD currents
with all Dirac structures $\Gamma$
has been obtained in Ref.~\cite{G:00}.

The QCD quantities $\bar{\Gamma}(mv,0)$, $Z_Q$, and $Z_q$
in the numerator of~(\ref{Match:main})
are calculated in terms of $\alpha_s^0=g_0^2/(4\pi)^{1-\varepsilon}$,
the bare coupling of the $n_f$-flavour QCD.
If there is another massive flavour, say charm,
they also contain its bare mass $m_{c0}$;
we re-express it via the on-shell mass $m_c$.
These quantities do not involve $\mu$.
The $\overline{\mbox{MS}}$ renormalization constant
$Z_j^{-1}(\mu)$ is expressed in terms of $\alpha_s^{(n_f)}(\mu)$.
The HQET quantities $\tilde{\Gamma}(0,0)$, $\tilde{Z}_Q$, and $\tilde{Z}_q$
in the denominator of~(\ref{Match:main})
are calculated via $\alpha_s^{0\prime}$ and $m_{c0}'$,
the bare coupling and the bare $c$-quark mass
in the $n_f'$-flavour QCD\footnote{To keep track of flavours in quark loops,
  we introduce the number $n_h$ of heavy flavours with mass $m$
  and the number $n_m$ of massive flavours with mass $m_c$,
  so that $n_f=n_l+n_m+n_h$ and $n_f'=n_l+n_m$.
  In reality, $n_h=1$ and $n_m=1$.
  If the $c$ quark is considered massless,
  we can also include it in the number of light flavours $n_l$,
  and set $n_m=0$.}
(we re-express $m_{c0}'$ via the on-shell mass $m_c$,
which is the same in both theories).
The $\overline{\mbox{MS}}$ renormalization constant
$\tilde{Z}_j^{-1}(\mu)$ is expressed in terms of $\alpha_s^{(n_f')}(\mu)$.
To combine all these quantities in~(\ref{Match:main}),
we re-express them via the coupling $\alpha_s^{(n_f')}(\mu)$
using the decoupling relation~\cite{CKS:98}
(an explicit expression for $\alpha_s^{(n_f)}(\mu)$
via $\alpha_s^{(n_f')}(\mu)$, including the necessary terms
with positive powers of $\varepsilon$,
is given in Eq.~(12) of Ref.~\cite{GMPS:08}).

There exists an exact relation~\cite{BG:95}
between the matching coefficients $C_1(\mu)$ and $C_{\rlap{\scriptsize/}v}(\mu)$.
Namely, the renormalized vector and scalar currents are related by
\begin{equation}
i \partial_\alpha j^\alpha = m(\mu) j(\mu)\,,
\label{Match:div}
\end{equation}
where $m(\mu)$ is the $\overline{\mbox{MS}}$ mass
of the heavy quark $Q$.
Taking the on-shell matrix element of this equality
between the heavy quark with $p=mv$ and the light quark with $k_q=0$
and re-expressing both QCD matrix elements via the matrix element
of the HQET current with $\Gamma=1$, we obtain
\begin{equation}
m C_{\rlap{\scriptsize/}v}(\mu) = m(\mu) C_1(\mu)\,.
\label{Match:mm}
\end{equation}
The ratio $m(\mu)/m$ has been calculated at three loops
in Refs.~\cite{CS:99,MR:00a}.
Comparing $C_{\rlap{\scriptsize/}v}(\mu)/C_1(\mu)$
with the analytical result~\cite{MR:00a} for $m(\mu)/m$
provides a strong check of our calculations.
For $m_c\neq0$, this ratio has been calculated in Ref.~\cite{BGSS:07}.


\section{Bare vertex functions}
\label{S:Calc}

\begin{figure}[h]
\parbox[b]{0.3\textwidth}{\begin{center}
    \epsfig{file=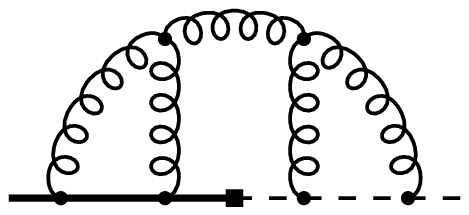,width=0.25\textwidth} 
  \end{center}}
\parbox[b]{0.3\textwidth}{\begin{center}
    \epsfig{file=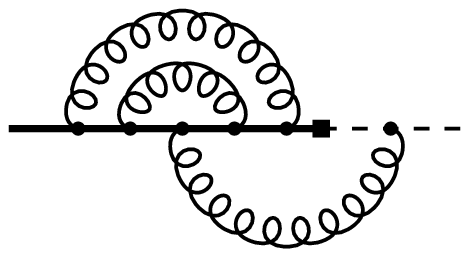,width=0.25\textwidth} 
  \end{center}}
 \parbox[b]{0.3\textwidth}{\begin{center}
    \epsfig{file=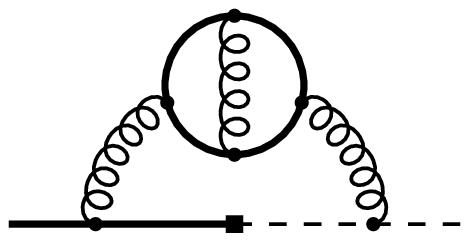,width=0.25\textwidth} 
  \end{center}}  \\[2ex]
\parbox[b]{0.3\textwidth}{\begin{center}
    \epsfig{file=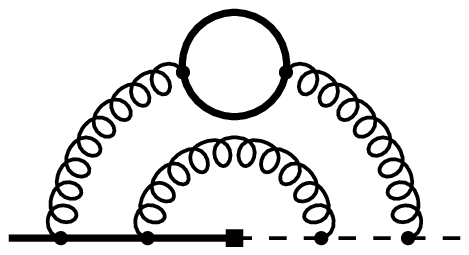,width=0.25\textwidth} 
  \end{center}}
\parbox[b]{0.3\textwidth}{\begin{center}
    \epsfig{file=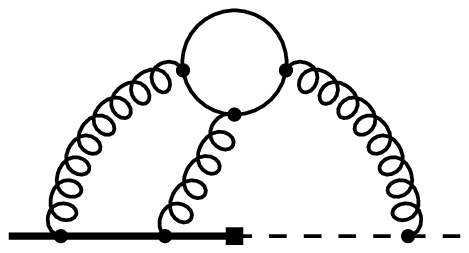,width=0.25\textwidth} 
  \end{center}}
\parbox[b]{0.3\textwidth}{\begin{center}
    \epsfig{file=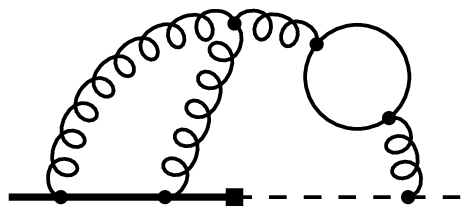,width=0.25\textwidth} 
  \end{center}} \\[2ex]
\parbox[b]{0.3\textwidth}{\begin{center}
    \epsfig{file=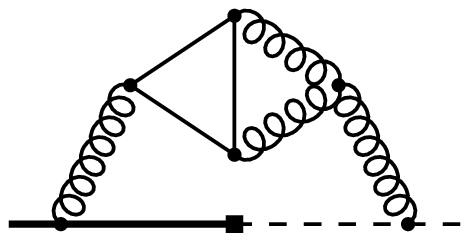,width=0.25\textwidth} 
  \end{center}}
\parbox[b]{0.3\textwidth}{\begin{center}
    \epsfig{file=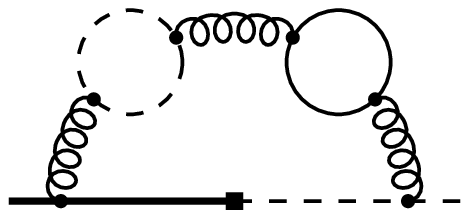,width=0.25\textwidth} 
  \end{center}}
\parbox[b]{0.3\textwidth}{\begin{center}
    \epsfig{file=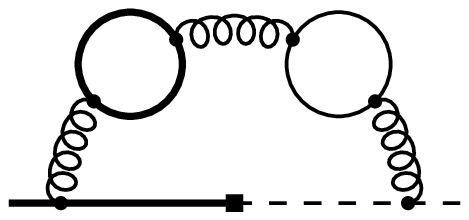,width=0.25\textwidth}
  \end{center}}
\caption{\label{fig:nm_diagrams} Sample Feynman diagrams for
  the matching coefficient. Thick, thin and dashed lines denote heavy,
  light and massless quarks, respectively; curly lines are gluons.
  The square denotes the vertex of the heavy--light current.}
\end{figure}

The calculation of $\bar{\Gamma}(mv,0)$ at $m_c=0$ is a single-scale problem.
The calculation is almost completely automated
and similar to Refs.~\cite{MMPS:07,GMPS:08}.
The Feynman diagrams were generated with \texttt{QGRAF}~\cite{N:91}
and classified into various topologies with the help of
\texttt{q2e} and \texttt{exp}~\cite{H:97}.
Some sample diagrams are shown in Fig.~\ref{fig:nm_diagrams}
(the first four diagrams).
Scalar Feynman integrals were reduced to master integrals using
integration-by-parts identities~\cite{IBP}. This was done in two independent ways:
using the \texttt{Form}~\cite{Vermaseren:2000nd} package \texttt{SHELL3}~\cite{MR:00}
and the \texttt{C++} program \texttt{Crusher}~\cite{PMDS}
implementing the Laporta algorithm~\cite{Laporta}.
The master integrals for the case of a massless charm quark are known
from Ref.~\cite{MR:00} (see also comments in Ref.~\cite{MMPS:07}).
As an independent check the automatic setup described in Ref.~\cite{MPSS:09}
has been used and the bare vertex functions have been checked numerically
with the help of \texttt{FIESTA}~\cite{Smirnov:2008py}.
The results for the bare vertex functions with all four Dirac structures $\Gamma$
can be found on the web page~\cite{web}.

Next we study the influence of a non-zero $c$-quark mass, $m_c<m$,
on the $b$-quark matching coefficients $C_\Gamma$.
Then $c$-quark loops exist both in the full QCD and in the $b$-quark HQET.
The full-theory quantities in the numerator of~(\ref{Match:main})
depend on the ratio of the on-shell quark masses,
\begin{equation}
  x = \frac{m_c}{m}\,.
\end{equation}
Some sample diagrams contributing to $\bar{\Gamma}(mv,0)$
are shown in Fig.~\ref{fig:nm_diagrams}
(all diagrams except the first four ones depend on $x$).
The HQET quantities in the denominator of~(\ref{Match:main})
contain a single scale $m_c$.

From the technical point of view the calculation is similar to Refs.~\cite{BGSS:07,BGSS:09}. 
We have used \texttt{Crusher}~\cite{PMDS} for the reduction.
The master integrals are known from Refs.~\cite{BGSS:07,BGSS:08},
see also Ref.~\cite{PHDStefan}.
Most of the needed terms of their $\varepsilon$ expansions
are known analytically, in terms of Harmonic Polylogarithms~\cite{RV:00,HPL}
(HPLs) of $x$ (the status of these expansions is summarized in Tables~1--4
of Ref.~\cite{BGSS:08}).
From the requirement of cancellation of $1/\varepsilon$ poles
in the matching coefficients (Sect.~\ref{S:Coefs}) we were able to find
exact analytical expressions (in terms of HPLs)
for the $\mathcal{O}(\varepsilon^0)$ terms of the master integrals 5.2 and 5.2a
(Fig.~8 in~\cite{BGSS:08});
formerly, they were known only as truncated series in $x$.
The corresponding entry in Table~3 of Ref.~\cite{BGSS:08} needs updating.
We do not present these long expressions here,
they can be found at~\cite{web}.

We have checked the $m_c$ dependent results by taking the limit $x\to0$
and reproducing the $n_l$ part of the results given in Sect.~\ref{S:Coefs}.
Another check is taking the limit $x \to 1$.
If we set $n_h=0$ and re-express the renormalized matching coefficients $C_\Gamma$
via $\alpha_s^{(n_l)}$, the results of Sect.~\ref{S:Coefs} with the substitution $n_m\to n_h$
are reproduced.
\section{Wave-function renormalization of massless quarks}
\label{S:Zq}

\begin{figure}
\begin{center}
\epsfig{file=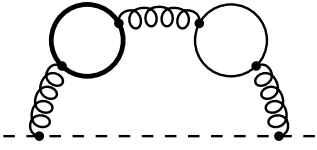,width=0.3\textwidth} 
\end{center}
\caption{\label{fig:zq}
The diagram which gives the coefficient of $n_m n_h$ in $Z_q$.
For the meaning of the lines see Fig.~\ref{fig:nm_diagrams}.}
\end{figure}

The last  ingredient of Eq. (\ref{Match:main}) which we had to calculate
is the on-shell wave-function renormalization constant $Z_q$ of a massless quark.
The result for \mbox{$m_c=0$} can be extracted from Ref.~\cite{CKS:98},
however, we have performed an independent calculation.
The $m_c$-dependent part is a new result.
The calculation is similar to that of Ref.~\cite{BGSS:07},
where $Z_Q$ has been calculated.
The integrals contributing to $Z_q$ reduce to tadpoles
when the mass of the incoming particle is set to zero.
There is only one new type of diagram which is shown in Fig.~\ref{fig:zq},
all others can be reduced to known results.

For completeness we give here the result for an arbitrary gauge group. It reads
\begin{eqnarray}
Z_q &{}={}& 1 + C_F T_F
\left(\frac{\alpha_s^0}{\pi}\Gamma(\varepsilon)\right)^2
\left(n_h m^{-4\varepsilon} + n_m m_c^{-4\varepsilon}\right)
\nonumber\\
&&\hphantom{{}+C_F T_F} \times
\frac{\varepsilon}{16}
\left(1
  - \frac{5}{6} \varepsilon
  + \frac{89}{36} \varepsilon^2
  + \mathcal{O}(\varepsilon^3)
\right)
\nonumber\\
&&{} + C_F T_F
\left(\frac{\alpha_s^0}{\pi}\Gamma(\varepsilon)\right)^3
\Bigl(n_h Z_1(n_h) m^{-6\varepsilon} + n_m Z_1(n_m) m_c^{-6\varepsilon}
\nonumber\\
&&\hphantom{{}+C_F T_F}
+ T_F n_h n_m (m m_c)^{-3\varepsilon} Z_2(m_c/m) \Bigr)
+ \mathcal{O}(\alpha_s^4)\,,
\end{eqnarray}
where $\alpha_s^0$ has the dimensionality $m^{2\varepsilon}$,
and the single-scale contributions are~\cite{CKS:98,GSS:06}
\begin{eqnarray}
Z_1(n) &{}={}&
C_F Z_F + C_A Z_A + T_F n_l Z_L + T_F n Z_H\,,
\nonumber\\
Z_F &{}={}& \frac{\varepsilon}{96}
\left[ 1 - \frac{3}{2} \varepsilon
  + \left(12 \zeta_3 + \frac{443}{12}\right) \varepsilon^2
\right]
+ \mathcal{O}(\varepsilon^4)\,,
\nonumber\\
Z_A &{}={}& \frac{1}{192} \biggl\{
1 + \frac{10}{3} \varepsilon + \frac{227}{9} \varepsilon^2
  - \left(16 \zeta_3 + \frac{1879}{54}\right) \varepsilon^3
\nonumber\\
&&{} - \xi
\left[ 1 - 3 \varepsilon + \frac{35}{3} \varepsilon^2
  + \left(8 \zeta_3 - \frac{407}{9}\right) \varepsilon^3
\right] \biggr\}
+ \mathcal{O}(\varepsilon^4)\,,
\nonumber\\
Z_L &{}={}& - \frac{\varepsilon}{72}
\left[1 - \frac{5}{6} \varepsilon + \frac{337}{36} \varepsilon^2\right]
+ \mathcal{O}(\varepsilon^4)\,,
\nonumber\\
Z_H &{}={}& - \frac{\varepsilon}{36}
\left[1 - \frac{5}{6} \varepsilon + \frac{151}{36} \varepsilon^2\right]
+ \mathcal{O}(\varepsilon^4)\,,
\end{eqnarray}
where $\xi=1-a_0$, $a_0$ is the bare gauge-fixing parameter.%
\footnote{The gauge-fixing term in the Lagrangian is
$-\left(\partial_\mu A_0^{a\mu}\right)^2/(2 a_0)$,
the free gluon propagator is
$-(i/k^2) \left(g_{\mu\nu} - \xi k_\mu k_\nu/k^2\right)$.}
The two-scale contribution (Fig.~\ref{fig:zq})
is given by~\cite{PHDStefan}
\begin{equation}
Z_2(x) = Z_2(x^{-1}) = 2 Z_H - \frac{\varepsilon^3}{12} \ln^2 x
+ \mathcal{O}(\varepsilon^4)\,.
\end{equation}

\section{Matching coefficients}
\label{S:Coefs}

In this Section, we present the results for the matching coefficients of the
different heavy--light currents for the colour group SU(3)
(results for a general colour group can be found at~\cite{web}).
For this purpose we decompose the coefficients as follows
\begin{eqnarray}
C_\Gamma(\mu) &{}={}& 1 + \frac{\alpha_s^{(n_f')}(m)}{\pi}\, C_\Gamma^{(1)}
+ \left(\frac{\alpha_s^{(n_f')}(m)}{\pi}\right)^2\, C_\Gamma^{(2)}(x)
\nonumber\\
&&{} + \left(\frac{\alpha_s^{(n_f')}(m)}{\pi}\right)^3\, C_\Gamma^{(3)}(x)
+ \mathcal{O}(\alpha_s^4)\,,
\\
C_\Gamma^{(2)}(x) &{}={}& C_\Gamma^G
+ C_\Gamma^H n_h + C_\Gamma^L n_l + C_\Gamma^M(x) n_m\,,
\nonumber\\
C_\Gamma^{(3)}(x) &{}={}& C_\Gamma^{GG} + C_\Gamma^{GH} n_h + C_\Gamma^{GL} n_l
+ C_\Gamma^{HH} n_h^2 + C_\Gamma^{HL} n_h n_l + C_\Gamma^{LL} n_l^2
\nonumber\\
&&{} + C_\Gamma^{GM}(x) n_m +C_\Gamma^{HM}(x) n_h n_m + C_\Gamma^{LM}(x) n_l n_m
+ C_\Gamma^{MM}(x) n_m^2\,,
\nonumber
\end{eqnarray}
where $\Gamma=1$, $\rlap/v$, $\gamma_\bot$, $\gamma_\bot\rlap/v$,
and $n_f'=n_l+n_m$ is the number of active flavours in HQET.
Furthermore, we use the abbreviation $\Lmu = \ln(\mu^2/m^2)$
($m$ is the on-shell $b$-quark mass).

We first present the results for $m_c = 0$. The individual contributions read
\begin{eqnarray}
  C^{(1)}_1 &{}={}&
  \frac{2}{3}
  + \frac{1}{2} \Lmu
  \,,\nonumber\\
  C^{G}_1 &{}={}&
  \frac{1843}{192}
  + \frac{11}{72} \pi^2
  + \frac{1}{18} \pi^2 \ln2
  - \frac{17}{36} \zeta_3
  + \left(
    \frac{527}{144}
    - \frac{7}{108} \pi^2
  \right) \Lmu
  - \frac{9}{16} \Lmu^2
  \,,\nonumber\\
  C^{H}_1 &{}={}&
    \frac{149}{216}
    - \frac{1}{18} \pi^2
    - \frac{5}{36} \Lmu
    + \frac{1}{12} \Lmu^2
  \,,\nonumber\\
  C^{L}_1 &{}={}&
    - \frac{95}{288}
    - \frac{1}{36} \pi^2
    - \frac{5}{72} \Lmu
    + \frac{1}{24} \Lmu^2
  \,,\nonumber\\
  C^{GG}_1 &{}={}&
  \frac{8765231}{62208}
  + \frac{235073}{46656} \pi^2
  + \frac{917}{324} \pi^2 \ln2
  - \frac{1}{81} \pi^2 \ln^22
  + \frac{3371}{1296} \zeta_3
  \nonumber\\
  &&{} + \frac{4733}{3888} \pi^2 \zeta_3
  - \frac{50039}{116640} \pi^4
  - \frac{28975}{2592} \zeta_5
  - \frac{4}{27} \ln^42
  - \frac{32}{9} a_4
  \nonumber\\
  &&{} + \left(
    \frac{46123}{1728}
    - \frac{25}{108} \pi^2
    + \frac{1}{36} \pi^2 \ln2
    + \frac{55}{144} \zeta_3
    - \frac{95}{1944} \pi^4
  \right) \Lmu
  \nonumber \\
  &&{} + \left(
    - \frac{605}{64}
    + \frac{7}{48} \pi^2
  \right) \Lmu^2
  + \frac{15}{16} \Lmu^3
  \,,\nonumber
\end{eqnarray}
\begin{eqnarray}
  C^{GH}_1 &{}={}&
  \frac{3349}{1944}
  - \frac{30917}{9720} \pi^2
  + \frac{443}{81} \pi^2 \ln2
  + \frac{1}{9} \pi^2 \ln^22
  - \frac{27845}{5184} \zeta_3
  + \frac{29}{96} \pi^2 \zeta_3
  \nonumber\\
  &&{} - \frac{19}{2430} \pi^4
  - \frac{45}{32} \zeta_5
  - \frac{1}{9} \ln^42
  - \frac{8}{3} a_4
  \nonumber\\
  &&{} + \left(
    - \frac{319}{432}
    - \frac{1}{36} \pi^2
    - \frac{5}{6} \zeta_3
  \right) \Lmu
  + \frac{211}{144} \Lmu^2
  - \frac{19}{72} \Lmu^3
  \,,\nonumber\\
  C^{GL}_1 &{}={}&
  - \frac{528353}{46656}
  - \frac{15553}{17496} \pi^2
  - \frac{25}{324} \pi^2 \ln2
  + \frac{1}{81} \pi^2 \ln^22
  - \frac{1591}{972} \zeta_3
  \nonumber\\
  &&{} + \frac{3281}{116640} \pi^4
  + \frac{1}{162} \ln^42
  + \frac{4}{27} a_4
  + \left(
    - \frac{7399}{5184}
    + \frac{11}{972} \pi^2
    - \frac{97}{216} \zeta_3
  \right) \Lmu
  \nonumber\\
  &&{}
  + \left(
    \frac{809}{864}
    - \frac{7}{648} \pi^2
  \right) \Lmu^2
  - \frac{19}{144} \Lmu^3
  \,,\nonumber\\
  C^{HH}_1 &{}={}&
  - \frac{4045}{11664}
  - \frac{1}{135} \pi^2
  + \frac{11}{27} \zeta_3
  - \frac{35}{1296} \Lmu
  - \frac{5}{216} \Lmu^2
  + \frac{1}{108} \Lmu^3
  \,,\nonumber\\
  C^{HL}_1 &{}={}&
  \frac{353}{5832}
  + \frac{1}{216} \pi^2
  - \frac{2}{27} \zeta_3
  - \frac{35}{648} \Lmu
  - \frac{5}{108} \Lmu^2
  + \frac{1}{54} \Lmu^3
  \,,\nonumber\\
  C^{LL}_1 &{}={}&
  \frac{6457}{46656}
  + \frac{13}{648} \pi^2
  + \frac{7}{108} \zeta_3
  - \frac{35}{2592} \Lmu
  - \frac{5}{432} \Lmu^2
  + \frac{1}{216} \Lmu^3
  \,,\\
  C^{(1)}_{\rlap{\scriptsize/}v} &{}={}&
  - \frac{2}{3}
  - \frac{1}{2} \Lmu
  \,,\nonumber\\
  C^{G}_{\rlap{\scriptsize/}v} &{}={}&
  - \frac{177}{64}
  - \frac{5}{72} \pi^2
  - \frac{1}{18} \pi^2 \ln2
  - \frac{11}{36} \zeta_3
  + \left(
    - \frac{79}{144}
    - \frac{7}{108} \pi^2
  \right) \Lmu
  + \frac{13}{16} \Lmu^2
  \,,\nonumber\\
  C^{H}_{\rlap{\scriptsize/}v} &{}={}&
    \frac{727}{432}
    - \frac{1}{6} \pi^2
  \,,\nonumber\\
  C^{L}_{\rlap{\scriptsize/}v} &{}={}&
    \frac{47}{288}
    + \frac{1}{36} \pi^2
    + \frac{5}{72} \Lmu
    - \frac{1}{24} \Lmu^2
  \,,\nonumber\\
  C^{GG}_{\rlap{\scriptsize/}v} &{}={}&
  - \frac{62575}{62208}
  - \frac{231253}{46656} \pi^2
  - \frac{517}{324} \pi^2 \ln2
  + \frac{20}{81} \pi^2 \ln^22
  + \frac{5645}{1296} \zeta_3
  \nonumber\\
  &&{} + \frac{2089}{486} \pi^2 \zeta_3
  - \frac{17347}{58320} \pi^4
  - \frac{49435}{2592} \zeta_5
  + \frac{11}{54} \ln^42
  + \frac{44}{9} a_4
  \nonumber\\
  &&{}
  + \bigg(
    \frac{115}{54}
    - \frac{121}{648} \pi^2
    + \frac{1}{36} \pi^2 \ln2
    + \frac{37}{48} \zeta_3
    - \frac{95}{1944} \pi^4
  \bigg) \Lmu
  \nonumber\\
  &&{}
  + \left(
    \frac{2257}{576}
    + \frac{91}{432} \pi^2
  \right) \Lmu^2
  - \frac{13}{8} \Lmu^3
  \,,\nonumber\\
  C^{GH}_{\rlap{\scriptsize/}v} &{}={}&
  \frac{2051}{96}
  - \frac{24583}{2430} \pi^2
  + \frac{361}{27} \pi^2 \ln2
  + \frac{10}{81} \pi^2 \ln^22
  - \frac{45869}{5184} \zeta_3
  + \frac{53}{96} \pi^2 \zeta_3
  \nonumber\\
  &&{}
  - \frac{1}{20} \pi^4
  - \frac{85}{32} \zeta_5
  - \frac{10}{81} \ln^42
  - \frac{80}{27} a_4
  + \left(
    - \frac{727}{864}
    + \frac{1}{12} \pi^2
  \right) \Lmu
  \,,\nonumber\\
  C^{GL}_{\rlap{\scriptsize/}v} &{}={}&
  \frac{24457}{46656}
  + \frac{5575}{8748} \pi^2
  + \frac{19}{324} \pi^2 \ln2
  - \frac{1}{81} \pi^2 \ln^22
  + \frac{3181}{1944} \zeta_3
  - \frac{379}{116640} \pi^4
  \nonumber\\
  &&{}
  - \frac{1}{162} \ln^42
  - \frac{4}{27} a_4
  + \left(
    - \frac{319}{5184}
    + \frac{11}{972} \pi^2
    + \frac{83}{216} \zeta_3
  \right) \Lmu
  \nonumber\\
  &&{}
  + \left(
    - \frac{469}{864}
    - \frac{7}{648} \pi^2
  \right) \Lmu^2
  + \frac{25}{144} \Lmu^3
  \,,\nonumber
\end{eqnarray}
\begin{eqnarray}
  C^{HH}_{\rlap{\scriptsize/}v} &{}={}&
  - \frac{5857}{7776}
  + \frac{1}{405} \pi^2
  + \frac{11}{18} \zeta_3
  \,,\nonumber\\
  C^{HL}_{\rlap{\scriptsize/}v} &{}={}&
  - \frac{193}{432}
  + \frac{29}{648} \pi^2
  \,,\nonumber\\
  C^{LL}_{\rlap{\scriptsize/}v} &{}={}&
  \frac{1751}{46656}
  - \frac{13}{648} \pi^2
  - \frac{7}{108} \zeta_3
  + \frac{35}{2592} \Lmu
  + \frac{5}{432} \Lmu^2
  - \frac{1}{216} \Lmu^3
  \,,\\
  C^{(1)}_{\gamma_\bot} &{}={}&
  - \frac{4}{3}
  - \frac{1}{2} \Lmu
  \,,\nonumber\\
  C^{G}_{\gamma_\bot} &{}={}&
  - \frac{14651}{1728}
  - \frac{125}{648} \pi^2
  - \frac{7}{54} \pi^2 \ln2
  - \frac{7}{36} \zeta_3
  + \left(
    - \frac{31}{144}
    - \frac{7}{108} \pi^2
  \right) \Lmu
  \nonumber\\
  &&{} + \frac{13}{16} \Lmu^2
  \,,\nonumber\\
  C^{H}_{\gamma_\bot} &{}={}&
    \frac{133}{144}
    - \frac{5}{54} \pi^2
  \,,\nonumber\\
C^{L}_{\gamma_\bot} &{}={}&
    \frac{445}{864}
    + \frac{1}{36} \pi^2
    + \frac{5}{72} \Lmu
    - \frac{1}{24} \Lmu^2
  \,,\nonumber\\
  C^{GG}_{\gamma_\bot} &{}={}&
  - \frac{5046967}{62208}
  - \frac{361033}{46656} \pi^2
  - \frac{1745}{324} \pi^2 \ln2
  + \frac{124}{243} \pi^2 \ln^22
  + \frac{3929}{1296} \zeta_3
  \nonumber\\
  &&{}
  + \frac{3463}{972} \pi^2 \zeta_3
  - \frac{461}{3888} \pi^4
  - \frac{43835}{2592} \zeta_5
  + \frac{163}{486} \ln^42
  + \frac{652}{81} a_4
  \nonumber\\
  &&{}
  + \left(
    \frac{301}{54}
    - \frac{53}{648} \pi^2
    + \frac{7}{108} \pi^2 \ln2
    + \frac{103}{144} \zeta_3
    - \frac{95}{1944} \pi^4
  \right) \Lmu
  \nonumber\\
  &&{}
  + \left(
    \frac{1945}{576}
    + \frac{91}{432} \pi^2
  \right) \Lmu^2
  - \frac{13}{8} \Lmu^3
  \,,\nonumber\\
  C^{GH}_{\gamma_\bot} &{}={}&
  \frac{4133}{288}
  - \frac{3385}{486} \pi^2
  + \frac{2069}{243} \pi^2 \ln2
  + \frac{26}{243} \pi^2 \ln^22
  - \frac{10445}{5184} \zeta_3
  \nonumber\\
  &&{}
  + \frac{35}{288} \pi^2 \zeta_3
  - \frac{233}{14580} \pi^4
  - \frac{35}{96} \zeta_5
  - \frac{26}{243} \ln^42
  - \frac{208}{81} a_4
  \nonumber\\
  &&{}
  + \left(
    - \frac{133}{288}
    + \frac{5}{108} \pi^2
  \right) \Lmu
  \,,\nonumber\\
  C^{GL}_{\gamma_\bot} &{}={}&
  \frac{455461}{46656}
  + \frac{4937}{4374} \pi^2
  + \frac{169}{972} \pi^2 \ln2
  - \frac{7}{243} \pi^2 \ln^22
  + \frac{5173}{1944} \zeta_3
  \nonumber\\
  &&{}
  - \frac{2963}{116640} \pi^4
  - \frac{7}{486} \ln^42
  - \frac{28}{81} a_4
  + \left(
    - \frac{1471}{5184}
    + \frac{11}{972} \pi^2
    + \frac{83}{216} \zeta_3
  \right) \Lmu
  \nonumber\\
  &&{}
  + \left(
    - \frac{445}{864}
    - \frac{7}{648} \pi^2
  \right) \Lmu^2
  + \frac{25}{144} \Lmu^3
  \,,\nonumber\\
  C^{HH}_{\gamma_\bot} &{}={}&
  - \frac{3641}{7776}
  + \frac{11}{1215} \pi^2
  + \frac{17}{54} \zeta_3
  \,,\nonumber\\
  C^{HL}_{\gamma_\bot} &{}={}&
  - \frac{2545}{3888}
  + \frac{127}{1944} \pi^2
  \,,\nonumber\\
  C^{LL}_{\gamma_\bot} &{}={}&
  - \frac{7993}{46656}
  - \frac{7}{216} \pi^2
  - \frac{7}{108} \zeta_3
  + \frac{35}{2592} \Lmu
  + \frac{5}{432} \Lmu^2
  - \frac{1}{216} \Lmu^3
  \,,\\
  C^{(1)}_{\gamma_\bot\rlap{\scriptsize/}v} &{}={}&
  - \frac{4}{3}
  - \frac{5}{6} \Lmu
  \,,\nonumber\\
  C^{G}_{\gamma_\bot\rlap{\scriptsize/}v} &{}={}&
  - \frac{20749}{1728}
  - \frac{7}{24} \pi^2
  - \frac{1}{6} \pi^2 \ln2
  - \frac{5}{36} \zeta_3
  + \left(
    - \frac{329}{144}
    - \frac{7}{108} \pi^2
  \right) \Lmu
  \nonumber\\
  &&{}
  + \frac{215}{144} \Lmu^2
  \,,\nonumber
\end{eqnarray}
\begin{eqnarray}
  C^{H}_{\gamma_\bot\rlap{\scriptsize/}v} &{}={}&
    \frac{809}{648}
    - \frac{7}{54} \pi^2
    + \frac{13}{108} \Lmu
    - \frac{1}{36} \Lmu^2
  \,,\nonumber\\
  C^{L}_{\gamma_\bot\rlap{\scriptsize/}v} &{}={}&
    \frac{1745}{2592}
    + \frac{5}{108} \pi^2
    + \frac{41}{216} \Lmu
    - \frac{5}{72} \Lmu^2
  \,,\nonumber\\
  C^{GG}_{\gamma_\bot\rlap{\scriptsize/}v} &{}={}&
  - \frac{21556403}{186624}
  - \frac{488167}{46656} \pi^2
  - \frac{757}{108} \pi^2 \ln2
  + \frac{142}{243} \pi^2 \ln^22
  + \frac{8357}{3888} \zeta_3
  \nonumber\\
  &&{}
  + \frac{16153}{3888} \pi^2 \zeta_3
  - \frac{2447}{23328} \pi^4
  - \frac{15925}{864} \zeta_5
  + \frac{113}{243} \ln^42
  + \frac{904}{81} a_4
  \nonumber\\
  &&{}
  + \left(
    \frac{7871}{1728}
    + \frac{7}{108} \pi^2
    + \frac{5}{36} \pi^2 \ln2
    + \frac{61}{48} \zeta_3
    - \frac{95}{1944} \pi^4
  \right) \Lmu
  \nonumber\\
  &&{}
  + \left(
    \frac{22177}{1728}
    + \frac{301}{1296} \pi^2
  \right) \Lmu^2
  - \frac{4085}{1296} \Lmu^3
  \,,\nonumber\\
  C^{GH}_{\gamma_\bot\rlap{\scriptsize/}v} &{}={}&
  \frac{125005}{5832}
  - \frac{268333}{29160} \pi^2
  + \frac{301}{27} \pi^2 \ln2
  + \frac{1}{9} \pi^2 \ln^22
  - \frac{22469}{5184} \zeta_3
  \nonumber\\
  &&{}
  + \frac{59}{288} \pi^2 \zeta_3
  - \frac{73}{2430} \pi^4
  - \frac{25}{32} \zeta_5
  - \frac{1}{9} \ln^42
  - \frac{8}{3} a_4
  \nonumber\\
  &&{}
  + \left(
    - \frac{1117}{3888}
    + \frac{35}{324} \pi^2
    + \frac{5}{18} \zeta_3
  \right) \Lmu
  - \frac{1225}{1296} \Lmu^2
  + \frac{1}{8} \Lmu^3
  \,,\nonumber\\
  C^{GL}_{\gamma_\bot\rlap{\scriptsize/}v} &{}={}&
  \frac{211705}{15552}
  + \frac{28133}{17496} \pi^2
  + \frac{71}{324} \pi^2 \ln2
  - \frac{1}{27} \pi^2 \ln^22
  + \frac{3347}{972} \zeta_3
  \nonumber\\
  &&{}
  - \frac{4183}{116640} \pi^4
  - \frac{1}{54} \ln^42
  - \frac{4}{9} a_4
  + \left(
    \frac{4091}{15552}
    - \frac{13}{972} \pi^2
    + \frac{143}{216} \zeta_3
  \right) \Lmu
  \nonumber\\
  &&{}
  + \left(
    - \frac{3845}{2592}
    - \frac{7}{648} \pi^2
  \right) \Lmu^2
  + \frac{5}{16} \Lmu^3
  \,,\nonumber\\
  C^{HH}_{\gamma_\bot\rlap{\scriptsize/}v} &{}={}&
  - \frac{21281}{34992}
  + \frac{1}{81} \pi^2
  + \frac{31}{81} \zeta_3
  + \frac{1}{144} \Lmu
  + \frac{13}{648} \Lmu^2
  - \frac{1}{324} \Lmu^3
  \,,\nonumber\\
  C^{HL}_{\gamma_\bot\rlap{\scriptsize/}v} &{}={}&
  - \frac{14567}{17496}
  + \frac{17}{216} \pi^2
  + \frac{2}{81} \zeta_3
  + \frac{1}{72} \Lmu
  + \frac{13}{324} \Lmu^2
  - \frac{1}{162} \Lmu^3
  \,,\nonumber\\
  C^{LL}_{\gamma_\bot\rlap{\scriptsize/}v} &{}={}&
  - \frac{29309}{139968}
  - \frac{89}{1944} \pi^2
  - \frac{35}{324} \zeta_3
  + \frac{53}{2592} \Lmu
  + \frac{41}{1296} \Lmu^2
  - \frac{5}{648} \Lmu^3
  \,,
\end{eqnarray}
where $a_4=\mathop{\mathrm{Li}}\nolimits_4\left(\frac{1}{2}\right)$.
The two-loop results, as well as the coefficients $C_\Gamma^{LL}$,
are known from Ref.~\cite{BG:95}; all remaining three-loop results are new.

It is instructive to re-write the matching coefficients for $m_c=0$ and $\mu=m$
via the leading $\beta$-function coefficient in HQET, $\beta_0'=11-\frac{2}{3}n_f'$.
We obtain
\begin{eqnarray}
C_1^{(2)}
&{}={}& 0.91 \beta_0' + 1.09 = 7.55 + 1.09 = 8.64\,,
\nonumber\\
C_{\rlap{\scriptsize/}v}^{(2)}
&{}={}& - 0.66 \beta_0' + 3.06 = - 5.47 + 3.06 = - 2.41\,,
\nonumber\\
C_{\gamma_\bot}^{(2)}
&{}={}& - 1.18 \beta_0' + 1.53 = - 9.87 + 1.53 = - 8.34\,,
\nonumber\\
C_{\gamma_\bot\rlap{\scriptsize/}v}^{(2)}
&{}={}& - 1.70 \beta_0' + 2.42 = - 14.13 + 2.42 = - 11.70
\label{nna2}
\end{eqnarray}
at two loops, and the following results at three loops:
\begin{eqnarray}
C_1^{(3)}
&{}={}& 0.93 \beta_0^{\prime2} + 9.04 \beta_0' - 38.16
= 64.74 + 75.34 - 38.16 = 101.92\,,
\nonumber\\
C_{\rlap{\scriptsize/}v}^{(3)}
&{}={}& - 0.54 \beta_0^{\prime2} - 1.29 \beta_0' + 29.74
= - 37.25 - 10.72 + 29.74 = - 18.23\,,
\nonumber\\
C_{\gamma_\bot}^{(3)}
&{}={}& - 1.28 \beta_0^{\prime2} - 5.56 \beta_0' + 45.34
= - 88.92 - 46.34 + 45.34 = - 89.92\,,
\nonumber\\
C_{\gamma_\bot\rlap{\scriptsize/}v}^{(3)}
&{}={}& - 1.78 \beta_0^{\prime2} - 7.63 \beta_0' + 63.22
= - 123.61 - 63.57 + 63.22
\nonumber\\
&{}={}& - 123.96\,.
\label{nna3}
\end{eqnarray}
A method to estimate higher loop contributions called naive nonabelianization
has been formulated in Ref.~\cite{BG:95}.
It is based on the fact that each polynomial in $n_f$
can be re-written as a polynomial in $\beta_0$.
Usually it is relatively simple to calculate
the term with the highest power of $n_f$.
This means that we know the coefficient of the leading power of $\beta_0$.
Neglecting subleading powers of $\beta_0$ we obtain an estimate of the full result.
The two-loop corrections to the matching coefficients~(\ref{nna2})
were among the examples confirming naive nonabelianization~\cite{BG:95}.
At three loops we see that this prescription reproduces the correct signs
and roughly the correct magnitude of the full results.
In all cases we observe a compensation between the $\mathcal{O}(\beta_0')$
and $\mathcal{O}(1)$ terms. In the case of $C_{\gamma_\bot}^{(3)}$ and
$C_{\gamma_\bot\rlap{\scriptsize/}v}^{(3)}$ this compensation is almost complete,
and naive nonabelianization works surprisingly well.

Since the expressions for the charm-mass dependence are quite involved
and not completely expressed in terms of HPLs of $x$
(the $\mathcal{O}(\varepsilon)$ terms of the master integrals 5.2, 5.2a, 5.3, 5.3a
are known analytically only as truncated series in $x$, see Table~3 in Ref.~\cite{BGSS:08}),
we refrain from listing them here. Instead, we present an 
expansion of our results to the second order in $x$. The results in
terms of the master integrals and expansions to higher orders can be obtained from the web
page~\cite{web}. 

Our results for the  $m_c$-dependent coefficients read
\begin{eqnarray}
C_1^{M}(x) &{}={}& C_1^{L}
+\frac{1}{8}\pi^2\, x
+\left(
     \ln x
     +\frac{1}{2}
\right) x^2
+\mathcal{O}(x^3)\,,
\nonumber\\
C_1^{GM}(x) &{}={}& C_1^{GL}
+\left(
     \frac{4361}{1296} \pi^2
     -\frac{119}{108} \pi^2 \ln 2
     +\frac{13}{216} \pi^3
     +\frac{1}{16} \pi^2 \Lmu
     -\frac{9}{8} \pi^2 \ln x
\right) x
\nonumber\\
&&{}
+\biggl[
     \frac{48493}{5184}
     +\frac{395}{432} \pi^2
     +\frac{7}{36} \pi^2 \ln 2
     +\frac{27}{16} \zeta_3
     -\pi^2 \zeta_3
     -\frac{49}{720} \pi^4
     -\frac{5}{2} \zeta_5
\nonumber\\
&&\hphantom{{}+\biggl[\biggr.}
     +\frac{1}{4} \Lmu
     +\left(
          \frac{6239}{432}
          +\frac{1}{4} \pi^2
          +\frac{15}{8} \zeta_3
          -\frac{19}{180} \pi^4
          +\frac{1}{2} \Lmu
     \right) \ln x
\nonumber\\
&&\hphantom{{}+\biggl[\biggr.}
     -\frac{37}{9} \ln^2 x
\biggr] x^2
+\mathcal{O}(x^3)\,,
\nonumber
\end{eqnarray}
\begin{eqnarray}
C_1^{HM}(x) &{}={}& C_1^{HL}
+\left(
     -\frac{517}{1350}
     +\frac{1}{18} \pi^2
     -\frac{1}{15} \ln x
\right) x^2
+ \mathcal{O}(x^3)\,,
\nonumber\\
C_1^{LM}(x) &{}={}& 2 C_1^{LL}
+\left(
    -\frac{7}{36} \pi^2
    +\frac{1}{6} \pi^2 \ln 2
    +\frac{1}{12} \pi^2 \ln x
\right) x
\nonumber\\
&&{}
+\left(
    -\frac{5}{18}
    -\frac{1}{18} \pi^2
    -\frac{2}{9} \ln x
    +\frac{1}{3} \ln^2 x
\right) x^2
+ \mathcal{O}(x^3)\,,
\nonumber\\
C_1^{MM}(x) &{}={}& C_1^{LL}
+\left(
     -\frac{1}{45} \pi^2
     +\frac{1}{12} \pi^2 \ln x
\right) x
\nonumber\\
&&{}
+\left(
     -\frac{19}{36}
     -\frac{1}{18} \pi^2
     -\frac{2}{9} \ln x
     +\frac{1}{3}\ln^2 x
\right) x^2
+\mathcal{O}(x^3)\,,
\\
C_{\rlap{\scriptsize/}v}^{M}(x) &{}={}& C_{\rlap{\scriptsize/}v}^{L}
-\frac{1}{24} \pi^2\, x
+\left(
     \frac{3}{2}
     +\ln x 
\right) x^2
+\mathcal{O}(x^3)\,,
\nonumber\\
C_{\rlap{\scriptsize/}v}^{GM}(x) &{}={}& C_{\rlap{\scriptsize/}v}^{GL}
+\biggl(
     -\frac{35333}{3888} \pi^2
     +\frac{4439}{324} \pi^2 \ln 2
     -\frac{13}{648} \pi^3
     +\frac{1}{48} \pi^2 \Lmu
\nonumber\\
&&\hphantom{C_{\rlap{\scriptsize/}v}^{GL}+\biggl(\biggr.}
     +\frac{17}{24} \pi^2 \ln x
\biggr) x
\nonumber\\
&&{}
+\biggl[
     \frac{86509}{5184}
     +\frac{863}{432} \pi^2
     +\frac{7}{36} \pi^2 \ln 2
     +\frac{115}{16} \zeta_3
     -\frac{7}{4} \pi^2 \zeta_3
     -\frac{49}{720} \pi^4
\nonumber\\
&&\hphantom{{}=\biggl[\biggr.}
     -5 \zeta_5
     -\frac{3}{4} \Lmu
     +\left(
          \frac{7391}{432}
          +\frac{7}{4} \pi^2
          +\frac{15}{8} \zeta_3
          -\frac{17}{90} \pi^4
          -\frac{1}{2} \Lmu
     \right) \ln x
\nonumber\\
&&\hphantom{{}=\biggl[\biggr.}
     -\frac{37}{9} \ln^2 x
\biggr] x^2
+\mathcal{O}(x^3)\,,
\nonumber\\
C_{\rlap{\scriptsize/}v}^{HM}(x) &{}={}& C_{\rlap{\scriptsize/}v}^{HL}
+\left(
     -\frac{817}{1350}
     +\frac{1}{18} \pi^2
     -\frac{1}{15} \ln x
\right) x^2
+\mathcal{O}(x^3)\,,
\nonumber\\
C_{\rlap{\scriptsize/}v}^{LM}(x) &{}={}& 2 C_{\rlap{\scriptsize/}v}^{LL}
+\left(\frac{7}{108} \pi^2
     -\frac{1}{18} \pi^2 \ln 2
     -\frac{1}{36} \pi^2 \ln x
\right) x
\nonumber\\
&&{}
+\left(
     -\frac{1}{2}
     -\frac{1}{18} \pi^2
     -\frac{2}{9} \ln x
     +\frac{1}{3} \ln^2 x
\right) x^2
+\mathcal{O}(x^3)\,,
\nonumber\\
C_{\rlap{\scriptsize/}v}^{MM}(x) &{}={}& C_{\rlap{\scriptsize/}v}^{LL}
+\left(
     \frac{1}{135} \pi^2
     -\frac{1}{36} \pi^2 \ln x
\right) x
\nonumber\\
&&{}
+\left(
     -\frac{3}{4}
     -\frac{1}{18} \pi^2
     -\frac{2}{9} \ln x
     +\frac{1}{3} \ln^2 x
\right) x^2 
+\mathcal{O}(x^3)\,,
\\
C_{\gamma_\bot}^{M}(x) &{}={}& C_{\gamma_\bot}^{L}
-\frac{1}{24} \pi^2\, x
+\left(
     \frac{3}{2}
     +\ln x
\right) x^2
+\mathcal{O}(x^3)\,,
\nonumber\\
C_{\gamma_\bot}^{GM}(x) &{}={}& C_{\gamma_\bot}^{GL}
+\biggl[
     -\frac{130327}{11664} \pi^2
     +\frac{14089}{972} \pi^2 \ln 2 
     -\frac{143}{1944} \pi^3
     +\frac{11}{144} \pi^2 \Lmu
\nonumber\\
&&\hphantom{C_{\gamma_\bot}^{GL}+\biggl[\biggr.}
     +\frac{409}{216} \pi^2 \ln x
\biggr] x
\nonumber
\end{eqnarray}
\begin{eqnarray}
&&{}
+\biggl[
     \frac{24217}{5184}
     +\frac{1517}{1296} \pi^2
     -\frac{7}{108} \pi^2 \ln 2
     +\frac{595}{144} \zeta_3
     -\frac{7}{12} \pi^2 \zeta_3
\nonumber\\
&&\hphantom{{}+\biggl[\biggr.}
     -\frac{19}{2160} \pi^4
     -\frac{5}{12} \Lmu
\nonumber\\
&&\hphantom{{}+\biggl[\biggr.}
     +\left(
          -\frac{79}{144}
          +\frac{595}{324} \pi^2
          -\frac{1}{8} \zeta_3
          -\frac{7}{135} \pi^4
          +\frac{1}{6} \Lmu
     \right) \ln x
\nonumber\\
&&\hphantom{{}+\biggl[\biggr.}
     +\frac{7}{9} \ln^2 x
\biggr] x^2
+\mathcal{O}(x^3)\,,
\nonumber\\
C_{\gamma_\bot}^{HM}(x) &{}={}& C_{\gamma_\bot}^{HL}
+\left(
     \frac{349}{4050}
     -\frac{1}{54} \pi^2
     -\frac{1}{15} \ln x
\right) x^2
+\mathcal{O}(x^3)\,,
\nonumber\\
C_{\gamma_\bot}^{LM}(x) &{}={}& 2 C_{\gamma_\bot}^{LL}
+\left(
     \frac{77}{324} \pi^2
     -\frac{11}{54} \pi^2 \ln 2
     -\frac{11}{108} \pi^2 \ln x
\right) x
\nonumber\\
&&{}
+\left(
     -\frac{19}{54}
     +\frac{1}{54} \pi^2
     +\frac{2}{27} \ln x
     -\frac{1}{9} \ln^2 x
\right) x^2
+\mathcal{O}(x^3)\,,
\nonumber\\
C_{\gamma_\bot}^{MM}(x) &{}={}& C_{\gamma_\bot}^{LL}
+\left(
     \frac{11}{405} \pi^2
     -\frac{11}{108} \pi^2 \ln x
\right) x
\nonumber\\
&&{}
+\left(
     -\frac{29}{108}
     +\frac{1}{54} \pi^2
     +\frac{2}{27} \ln x
     -\frac{1}{9} \ln^2 x
\right) x^2 
+\mathcal{O}(x^3)\,,
\\
C_{\gamma_\bot\rlap{\scriptsize/}v}^{M}(x) &{}={}& C_{\gamma_\bot\rlap{\scriptsize/}v}^{L}
-\frac{5}{24} \pi^2\, x
+\left(
     \frac{7}{6}
     -\frac{1}{3} \ln x
\right) x^2
+\mathcal{O}(x^3)\,,
\nonumber\\
C_{\gamma_\bot\rlap{\scriptsize/}v}^{GM}(x) &{}={}& C_{\gamma_\bot\rlap{\scriptsize/}v}^{GL}
+\biggl(
     -\frac{59221}{3888} \pi^2
     +\frac{6295}{324} \pi^2 \ln 2
     -\frac{65}{648} \pi^3
     +\frac{25}{144} \pi^2 \Lmu
\nonumber\\
&&\hphantom{C_{\gamma_\bot\rlap{\scriptsize/}v}^{GL}+\biggl(\biggr.}
     +\frac{541}{216} \pi^2 \ln x
\biggr) x
\nonumber\\
&&{}
+\biggl[
     \frac{35653}{5184}
     +\frac{631}{432} \pi^2
     -\frac{7}{108} \pi^2 \ln 2
     +\frac{313}{48} \zeta_3
     -\frac{5}{6} \pi^2 \zeta_3
     -\frac{19}{2160} \pi^4
\nonumber\\
&&\hphantom{{}+\biggl[\biggr.}
     -\frac{5}{6} \zeta_5
      -\frac{35}{36} \Lmu
\nonumber\\
&&\hphantom{{}+\biggl[\biggr.}
     +\left(
          \frac{575}{432}
          +\frac{9}{4} \pi^2
          -\frac{1}{8} \zeta_3
          -\frac{43}{540} \pi^4
          +\frac{5}{18} \Lmu
     \right) \ln x
\nonumber\\
&&\hphantom{{}+\biggl[\biggr.}
     +\frac{1}{2} \ln^2 x
\biggr] x^2
+\mathcal{O}(x^3)\,,
\nonumber\\
C_{\gamma_\bot\rlap{\scriptsize/}v}^{HM}(x) &{}={}& C_{\gamma_\bot\rlap{\scriptsize/}v}^{HL}
+\left(
     \frac{49}{4050}
     -\frac{1}{54} \pi^2
     -\frac{1}{15} \ln x
\right) x^2
+\mathcal{O}(x^3)\,,
\nonumber\\
C_{\gamma_\bot\rlap{\scriptsize/}v}^{LM}(x) &{}={}& 2 C_{\gamma_\bot\rlap{\scriptsize/}v}^{LL}
+\left(
     \frac{35}{108} \pi^2
     -\frac{5}{18} \pi^2 \ln 2
     -\frac{5}{36} \pi^2 \ln x
\right) x
\nonumber\\
&&{}
+\left(
     -\frac{23}{54}
     +\frac{1}{54} \pi^2
     +\frac{2}{27} \ln x
     -\frac{1}{9} \ln^2 x
\right) x^2
+\mathcal{O}(x^3)\,,
\nonumber\\
C_{\gamma_\bot\rlap{\scriptsize/}v}^{MM}(x) &{}={}& C_{\gamma_\bot\rlap{\scriptsize/}v}^{LL}
+\left(
     \frac{1}{27} \pi^2
     -\frac{5}{36} \pi^2 \ln x
\right) x
\nonumber\\
&&{}
+\left(
     -\frac{37}{108}
     +\frac{1}{54} \pi^2
     +\frac{2}{27} \ln x
     -\frac{1}{9} \ln^2 x
\right) x^2
+\mathcal{O}(x^3)\,.
\end{eqnarray}


\section{Meson matrix elements}
\label{S:Meson}

We are now in the position to apply our results to the matrix elements
between a $B$ or $B^*$ meson with momentum $p$ and the vacuum.
They are defined through
\begin{eqnarray}
{\langle}0| \left(\bar{q} \gamma_5^{\mbox{\scriptsize AC}} Q\right)_\mu |B{\rangle} &{}={}&
- i m_B f_B^P(\mu)\,,
\nonumber\\
{\langle}0| \bar{q} \gamma^\alpha \gamma_5^{\mbox{\scriptsize AC}} Q |B{\rangle} &{}={}&
i f_B p^\alpha\,,
\nonumber\\
{\langle}0| \bar{q} \gamma^\alpha Q |B^*{\rangle} &{}={}&
i m_{B^*} f_{B^*} e^\alpha\,,
\nonumber\\
{\langle}0| \left(\bar{q} \sigma^{\alpha\beta} Q\right)_\mu |B^*{\rangle} &{}={}&
f_{B^*}^{T}(\mu) (p^\alpha e^\beta - p^\beta e^\alpha)\,,
\label{Meson:matel}
\end{eqnarray}
where $e^\alpha$ is the $B^*$ polarization vector.
The corresponding HQET matrix elements (at $m\to\infty$) in the $v$ rest frame are
\begin{eqnarray}
{\langle}0| \left(\bar{q}\gamma_5^{\mbox{\scriptsize AC}} Q_v\right)_\mu
|B(\vec{k}\,){\rangle}\strut_{\mbox{\scriptsize nr}} &{}={}&
- i F(\mu)\,,
\nonumber\\
{\langle}0| \left(\bar{q} \vec{\gamma} Q_v\right)_\mu
|B^*(\vec{k}\,){\rangle}\strut_{\mbox{\scriptsize nr}} &{}={}&
i F(\mu) \vec{e}\,,
\label{Meson:HQET}
\end{eqnarray}
where the single-meson states are normalized by the non-relativistic condition
\[
\strut_{\mbox{\scriptsize nr}}{\langle}B(\vec{k}\,')|B(\vec{k}\,){\rangle}\strut_{\mbox{\scriptsize nr}}
= (2\pi)^3 \delta(\vec{k}\,'-\vec{k}\,)\,.
\]
We also remind the reader that $\bar{q} \Gamma \rlap/v Q_v = \bar{q} \Gamma Q_v$,
so that there are only two currents.
These two matrix elements are characterized by a single hadronic parameter $F(\mu)$
due to the heavy-quark spin symmetry.
From Eq.~(\ref{Match:div}) we have~\cite{BG:95}
\begin{equation}
\frac{f_B^P(\mu)}{f_B} = \frac{m_B}{m(\mu)}\,.
\label{Meson:mm}
\end{equation}
Here $m_B=m+\bar{\Lambda} + \mathcal{O}(\Lambda_{\mbox{\scriptsize QCD}}^2/m)$
where $\bar{\Lambda}$ is the residual energy of the ground-state $B$ meson
in the limit $m\to\infty$.
Neglecting $1/m$ corrections, we see that this equation coincides with~(\ref{Match:mm}).
We have checked that our results agree with the known formulas
for $m(\mu)/m$ at $m_c=0$~\cite{MR:00a}
and with $m_c$ corrections taken into account~\cite{BGSS:07}.

Using the expressions of Section~\ref{S:Coefs} we find the following
results for the ratios of decay constants. Again, we present an
expansion to the second order in $x$ for 
the charm mass dependent terms. For the numerical evaluations we use an
expansion to the eighth order in $x$ and the values $\alpha_s^{(4)}(m_b) =
0.2163$ and $x=0.3$.
\begin{eqnarray}
\frac{f_{B^*}}{f_B} &{}={}&
\frac{C_{\gamma_\bot}(m_b)}{C_{\rlap{\scriptsize/}v}(m_b)}
+ \mathcal{O}\left(\frac{\Lambda_{\mbox{\scriptsize QCD}}}{m_b}\right)
= 1 - \frac{2}{3} \frac{\alpha_s^{(4)}(m_b)}{\pi}
\nonumber\\
&&{} + \Bigl[ -7.749 - 0.028 n_h + 0.352 n_l
\nonumber\\
&&{} + \bigl(0.352 - 1.097 x -0.667 x^2 - 1.333 x^2 \ln x\bigr) n_m
\Bigr]
\left(\frac{\alpha_s^{(4)}(m_b)}{\pi}\right)^2
\nonumber\\
&&{}
+ \Bigl[ - 129.211 - 0.198 n_h + 14.294 n_l
  - 0.006 n_h^2 - 0.005 n_h n_l
\nonumber\\
&&{} - 0.331 n_l^2
  + \bigl( 14.294 - 17.816 x + 11.697 x \ln x
\nonumber\\
&&\qquad{} - 0.273 x^2 -6.082 x^2 \ln x + 4.889 x^2 \ln^2 x \bigr) n_m
\nonumber\\
&&{}
  + \bigl( -0.005 - 0.040 x^2 \bigr) n_h n_m
\nonumber\\
&&{}
  + \bigl( -0.661 + 0.692 x -0.731 x \ln x
\nonumber\\
&&\qquad{} + 0.879 x^2 + 0.296 x^2 \ln x -0.444 x^2 \ln^2 x \bigr) n_l n_m
\nonumber\\
&&{}
  + \bigl( - 0.331 + 0.195 x - 0.731 x \ln x
\nonumber\\
&&\qquad{} + 1.213 x^2 + 0.296 x^2 \ln x - 0.444 x^2 \ln^2 x \bigr) n_m^2
\Bigr]
\left(\frac{\alpha_s^{(4)}(m_b)}{\pi}\right)^3
\nonumber\\
&&{} + \mathcal{O}\left(\alpha_s^4,\frac{\Lambda_{\mbox{\scriptsize QCD}}}{m_b}\right)
\nonumber\\
&{}={}& 1 - \frac{2}{3} \frac{\alpha_s^{(4)}(m_b)}{\pi}
- (6.370 + 0.189) \left(\frac{\alpha_s^{(4)}(m_b)}{\pi}\right)^2
\nonumber\\
&&{}
- (77.549 + 6.575) \left(\frac{\alpha_s^{(4)}(m_b)}{\pi}\right)^3
+ \mathcal{O}\left(\alpha_s^4,\frac{\Lambda_{\mbox{\scriptsize QCD}}}{m_b}\right)
\nonumber\\
&{}={}& 1 - 0.046 - (0.030 + 0.001) - (0.025 + 0.002)
\nonumber\\
&{}={}& 0.899 - 0.003 = 0.896
+ \mathcal{O}\left(\alpha_s^4,\Lambda_{\mbox{\scriptsize QCD}}/m_b\right)\,.
\end{eqnarray}
In the second line from the bottom the corrections from tree level, first, second and
third order in $\alpha_s$ are given separately. Also the contributions stemming from
the finite charm mass are separated (the second number in the parentheses).
In the first part of the last line, the $m_c$ correction is also separated.
Power corrections $\mathcal{O}(\Lambda_{\mbox{\scriptsize QCD}}/m_b)$
are discussed in Refs.~\cite{N:92,CGM:03} and amount to several per
cent.

For the second ratio we obtain for $\mu=m_b$
\begin{eqnarray}
\frac{f_{B^*}^T(m_b)}{f_{B^*}} &{}={}&
\frac{C_{\rlap{\scriptsize/}v\gamma_\bot}(m_b)}{C_{\gamma_\bot}(m_b)}
+ \mathcal{O}\left(\frac{\Lambda_{\mbox{\scriptsize QCD}}}{m_b}\right)
\nonumber
\end{eqnarray}
\begin{eqnarray}
&{}={}& 1 +
\Bigl[ - 4.690 - 0.041 n_h + 0.341 n_l
\nonumber\\
&&{} + \bigl(0.341 - 0.548 x + 0.333 x^2 \bigr) n_m
\Bigr]
\left(\frac{\alpha_s^{(4)}(m_b)}{\pi}\right)^2
\nonumber\\
&&{}
+ \Bigl[ - 70.923 - 0.666 n_h + 9.175 n_l - 0.026 n_h^2 - 0.016 n_h n_l
\nonumber\\
&&{} - 0.222 n_l^2
+ \bigl(9.175 - 7.859 x + 6.031 x \ln x
\nonumber\\
&&\qquad{} + 4.555 x^2 + 3.256 x^2 \ln x - 0.278 x^2 \ln^2 x \bigr) n_m
\nonumber\\
&&{} + \bigl( - 0.016 - 0.074 x^2 \bigr) n_h n_m
\nonumber\\
&&{} + \bigl( - 0.444 + 0.346 x - 0.366 x \ln x - 0.074 x^2 \bigr) n_l n_m
\nonumber\\
&&{} + \bigl( - 0.222 + 0.097 x -0.366 x \ln x - 0.074 x^2 \bigr) n_m^2
\Bigr]
\left(\frac{\alpha_s^{(4)}(m_b)}{\pi}\right)^3
\nonumber\\
&&{} + \mathcal{O}\left(\alpha_s^4,\frac{\Lambda_{\mbox{\scriptsize QCD}}}{m_b}\right)
\nonumber\\
&{}={}& 1 - (3.367 + 0.142) \left(\frac{\alpha_s^{(4)}(m_b)}{\pi}\right)^2
\nonumber\\
&&{} - (38.530 + 3.973) \left(\frac{\alpha_s^{(4)}(m_b)}{\pi}\right)^3
+ \mathcal{O}\left(\alpha_s^4,\frac{\Lambda_{\mbox{\scriptsize QCD}}}{m_b}\right)
\nonumber\\
&{}={}& 1 - (0.016 + 0.001) - (0.013 + 0.001) \nonumber\\
&{}={}& 0.971 - 0.002 = 0.969
+ \mathcal{O}\left(\alpha_s^4,\Lambda_{\mbox{\scriptsize QCD}}/m_b\right)\,.
\end{eqnarray}

The coefficients of these perturbative series (at $m_c=0$ and $\mu=m_b$)
can be re-written via $\beta_0'$:
\begin{eqnarray}
\left(\frac{f_{B^*}}{f_B}\right)^{(2)} &{}={}&
- 0.53 \beta_0' - 1.97 = - 4.40 - 1.97 = - 6.37\,,
\nonumber\\
\left(\frac{f_{B^*}^T(m_b)}{f_{B^*}}\right)^{(2)} &{}={}&
- 0.51 \beta_0' + 0.89 = - 4.26 + 0.89 = - 3.37\,,
\nonumber\\
\left(\frac{f_{B^*}}{f_B}\right)^{(3)} &{}={}&
- 0.74 \beta_0^{\prime2} - 5.06 \beta_0' + 16.33
\nonumber\\
&{}={}& - 51.67 - 42.21 + 16.33 = - 77.55\,,
\nonumber\\
\left(\frac{f_{B^*}^T(m_b)}{f_{B^*}}\right)^{(3)} &{}={}&
- 0.50 \beta_0^{\prime2} - 2.75 \beta_0' + 19.07
\nonumber\\
&{}={}& - 34.69 - 22.91 + 19.07 = - 38.53\,.
\end{eqnarray}
Again, naive nonabelianization~\cite{BG:95} works quite well
for these ratios predicting the correct sign and order of magnitude.

Asymptotics of the perturbative coefficients for the matching coefficients
at a large number of loops $l\gg1$ have been investigated in Ref.~\cite{CGM:03}
in a model-independent way.
The results contain three unknown normalization constants $N_{0,1,2}\sim1$.
Let us in the following assume that the number of loops $l=3$
is much larger than one and compare our results with Ref.~\cite{CGM:03}.
The asymptotics of the perturbative coefficients for $f_{B^*}/f_B$
contain $N_0$ and $N_2$ (see~(5.6) in Ref.~\cite{CGM:03});
in the case of $m/\hat{m}$ it contains only $N_0$
(see~(5.9) in Ref.~\cite{CGM:03})%
\footnote{Note that for convenience the choice $\mu=m e^{-5/6}$
has been adopted in Ref.~\cite{CGM:03};
$\hat{m}$ is the renormalization-group invariant mass
(the exact definition used here is given in the unnumbered formula
after~(3.8) in Ref.~\cite{CGM:03}).}:
\begin{eqnarray}
\left(\frac{f_{B^*}}{f_B}\right)^{(n+1)}_{L=-5/3} &{}={}&
- \frac{14}{27} \left\{ 1 + \mathcal{O}\left(\frac{1}{n}\right)
+ \frac{2}{7} \left(\frac{50}{3} n\right)^{-9/25}
\left[1 + \mathcal{O}\left(\frac{1}{n}\right) \right]
\frac{N_2}{N_0} \right\}
\nonumber\\
&&{}\times\left(\frac{m}{\hat{m}}\right)^{(n+1)}_{L=-5/3}\,.
\label{asympt}
\end{eqnarray}
The coefficient of $N_2/N_0$ is about $0.08$ at $n=2$,
and it seems reasonable to neglect this contribution.
Neglecting also $1/n$ corrections, we obtain~\cite{CGM:03}
\begin{equation}
\left(\frac{f_{B^*}}{f_B}\right)^{(3)}_{L=-5/3} =
- \frac{14}{27} \cdot 56.37 = - 29.23\,.
\label{predict}
\end{equation}
Our exact result
\begin{equation}
\left(\frac{f_{B^*}}{f_B}\right)^{(3)}_{L=-5/3} = -37.787
\label{exact}
\end{equation}
agrees with this prediction reasonably well,
thus confirming the simple relation~(5.14) in Ref.~\cite{CGM:03}.
However, $1/n$ corrections are large and tend to break this agreement.
It is natural to expect that $1/n^2$ (and higher) corrections
are also substantial at $n=2$.
\section{Conclusion}
\label{S:Conc}

We have calculated the N$^3$LO corrections to the matching coefficients of
heavy--light currents in HQET. Our result takes into account effects due to
the mass of the charm quark. Strictly speaking, our results should be used
together with the N$^3$LO $\beta$-function~\cite{vanRitbergen:1997va,Czakon:2004bu}
and anomalous dimensions of both the QCD currents and the HQET one.
The four-loop anomalous dimensions are known for some of the QCD currents
(for $\Gamma=\rlap/v$, $\gamma_\bot$ the anomalous dimension is exactly zero;
for $\Gamma=1$ it is just the anomalous dimension of the $\overline{\mbox{MS}}$
mass~\cite{Chetyrkin:1997dh,Vermaseren:1997fq}
with a minus sign). However, the four-loop anomalous dimension of the HQET current
is not known (this anomalous dimension does not depend on the Dirac structure
$\Gamma$).%
\footnote{If a matrix element of the HQET current is extracted (from lattice
simulations or sum rules) at a low scale $\mu<m_c$ (without dynamic $c$ quarks),
then, in addition to HQET evolution without $c$ quark ($\mu<m_c$) and with it
($\mu>m_c$), one has to know the decoupling relation at $\mu=m_c$.
It is known with three-loop (N$^3$LO) accuracy~\cite{GSS:06}.}
The effect of this unknown anomalous dimension cancels in ratios
of $B$-meson decay constants, $f_{B^*}/f_B$ and $f_{B^*}^T(\mu)/f_{B^*}$,
discussed in Sect.~\ref{S:Meson}.

All the previous experience shows that
contributions from anomalous dimensions are numerically much smaller than from
matching coefficients. Matching coefficients have renormalon singularities at
the Borel parameter $u=1/2$, this is the position closest to the origin out
of all possible ones; this means that the factorial growth of their perturbative
coefficients is fastest among all possible variants. On the other hand, it is generally believed
that anomalous dimensions have no renormalon singularities, and their perturbative series
have finite convergence radii.

Only $f_B$ has been measured experimentally~\cite{exp}.
Our results can be used for predicting the $B^*$ decay constants. We find that
the perturbative series for $f_{B^*}/f_B$ and $f_{B^*}^T/f_{B^*}$
converge very slowly at best. The effects due to the
charm-quark mass are small and of the order of $10^{-3}$.

The matching coefficients $C_\Gamma$ can be used for extraction of $B$ (and $B^*$)
decay constants from lattice HQET simulations (see Ref.~\cite{lattice:09} for
recent reviews), or from HQET sum rules.
\section*{Acknowledgements}

We are grateful to T.~van~Ritbergen
for the package \texttt{SHELL3}~\cite{MR:00}.
We thank Rainer Sommer for communication.
This work was supported by the DFG through SFB/TR9
and the Graduiertenkolleg ``Hochenergiephysik und Teilchenastrophysik'',
by NSERC and the Alberta Ingenuity Fund.
The Feynman diagrams were drawn with the help of
\texttt{Axodraw}~\cite{Vermaseren:1994je}
and \texttt{JaxoDraw}~\cite{Binosi:2003yf}.



\end{document}